\documentclass[twocolumn]{jpsj2}

\usepackage{times}
\usepackage{graphicx}

\title{Effective Crystalline Electric Field Potential
in a $\mib{j}$-$\mib{j}$ Coupling Scheme}

\author{Takashi {\sc Hotta}$^1$ and Hisatomo {\sc Harima}$^2$}

\inst{$^1$Advanced Science Research Center,
Japan Atomic Energy Agency, Tokai, Ibaraki 319-1195 \\
$^2$Department of Physics, Faculty of Science, Kobe University,
Kobe 657-8501}

\recdate{\today}

\abst{
We propose an effective model on the basis of a $j$-$j$ coupling scheme
to describe local $f$-electron states for realistic values of
Coulomb interaction $U$ and spin-orbit coupling $\lambda$,
for future development of microscopic theory of magnetism and
superconductivity in $f^n$-electron systems,
where $n$ is the number of local $f$ electrons.
The effective model is systematically constructed by including
the effect of a crystalline electric field (CEF) potential
in the perturbation expansion in terms of $1/\lambda$.
In this paper, we collect all the terms up to the first order of $1/\lambda$.
Solving the effective model, we show the results of the CEF states
for each case of $n$=2$\sim$5 with $O_{\rm h}$ symmetry
in comparison with those of the Stevens Hamiltonian for the weak CEF.
In particular, we carefully discuss the CEF energy levels
in an intermediate coupling region with $\lambda/U$
in the order of 0.1 corresponding to actual $f$-electron materials
between the $LS$ and $j$-$j$ coupling schemes.
Note that the relevant energy scale of $U$ is the Hund's rule interaction.
It is found that the CEF energy levels in the intermediate coupling region
can be quantitatively reproduced by our modified $j$-$j$ coupling scheme,
when we correctly take into account the corrections in the order of $1/\lambda$
in addition to the CEF terms and Coulomb interactions
which remain in the limit of $\lambda$=$\infty$.
As an application of the modified $j$-$j$ coupling scheme,
we discuss the CEF energy levels of filled skutterudites
with $T_{\rm h}$ symmetry.
}

\kword{Crystalline electric field, spin-orbit interaction,
$j$-$j$ coupling scheme, $LS$ coupling scheme, filled skutterudites}

\begin{document}

\maketitle

%
%
\section{Introduction}

Theory of crystalline electric field (CEF) has been developed so far
on the basis of an $LS$ coupling scheme.\cite{Stevens,LLW,Hutchings}
Nowadays the CEF analysis is reduced to an automatic procedure,
since we can simply refer the Hutchings table for CEF parameters,
\cite{Hutchings}
$B_p^q$, depending on the value of total angular momentum $J$
of relevant rare-earth or actinide ion.
For instance, for Ce$^{3+}$ ion with one $f$ electron,
due to a spin-orbit coupling, the ground state is
included in the $J$=5/2 sextet.
By consulting the Hutchings table for $J$=5/2,\cite{Hutchings}
we easily find the matrix elements for CEF potential
among six states of $J$=5/2.
For a cubic system with $O_{\rm h}$ symmetry,
after some algebraic calculations, we immediately understand
that the $J$=5/2 sextet is split into $\Gamma_{7}^{-}$ doublet
and $\Gamma_{8}^{-}$ quartet.

For the purpose to fit the experimental results
on $f$-electron materials in a high-temperature region,
the CEF analysis using the Hutchings table is quite convenient,
since it is not necessary to have deep knowledge on
the origin of $B_p^q$, which is given by the sum of
electrostatic potentials from the ligand anions
surrounding the rare-earth or actinide ion.
In fact, for the case of Pr$^{3+}$ ion including two $f$ electrons,
it is enough to consider the CEF effect for $J$=4 nontet,
which is obtained after the consideration of the Hund's rules and
the spin-orbit interaction in the $LS$ coupling scheme.
Thus, the competition between CEF effect and Coulomb interactions
does not seem to appear in the actual calculation.
This fact reduces our task in the CEF analysis.

If $f$ electrons are perfectly localized,
we can satisfy with the above treatment and
further improvement on the CEF theory may not be required.
However, when $f$-electron properties are gradually changed from
localized to itinerant nature, there always appear rich phenomena
including exotic magnetism and unconventional superconductivity
in $f$-electron systems.
In order to describe the low-energy excitation of $f$ electrons,
we should improve the CEF theory so as to be compatible with
the $f$-electron Bloch state.
For such a purpose, the $LS$ coupling scheme is inconvenient,
since the CEF potential is applied to the multi-$f$-electron state
which is firmly constructed from the Hund's rules and
the spin-orbit interaction.
Rather, a $j$-$j$ coupling scheme is more convenient,
since individual $f$-electron states are first defined
under the effect of CEF potential.
Then, we construct the multi-$f$-electron state
by considering the effect of Coulomb interactions,
using standard quantum-field theoretical techniques.
In contrast, in the $LS$ coupling scheme,
we cannot use such standard techniques,
since Wick's theorem does not hold.

In order to describe the $f$-electron system by the Bloch state,
it is standard to exploit the relativistic electron band theory,
which is based on the $j$-$j$ coupling scheme.
In this sense, it is quite natural to construct
an $f$-electron model on the basis of the $j$-$j$ coupling scheme.
By keeping the same $f$-electron basis in a simple tight-binding model
as that of the relativistic band-structure calculation,
we can determine the hopping amplitudes in the tight-binding model
to reproduce the Fermi-surface structure
of the relativistic band-structure calculation result.
After that, we attempt to include electron correlations
between $f$ electrons in the tight-binding model.

On the above background, it has been recently proposed
to construct a microscopic model for $f$-electron compounds
on the basis of the $j$-$j$ coupling scheme,
\cite{Hotta1a,Hotta1b}
by applying a tight-binding approximation for kinetic part of $f$ electrons.
Then, microscopic theories have been developed
for the understanding of novel magnetism
\cite{Hotta2a,Onishi1,Hotta2b,Hotta3}
and unconventional superconductivity
\cite{Takimoto1,Takimoto2,Takimoto3,Hotta4,Kubo1a,Kubo1b}
in $f$-electron systems.
In addition, it is also possible to study complex multipole
phenomena from a microscopic viewpoint
by using the same $f$-electron model.
\cite{Kubo2a,Kubo2b,Kubo2c,Hotta5a,Hotta5b,Onishi2}

However, in the standard $j$-$j$ coupling scheme
in which $j$=7/2 states are simply discarded,
only second- and fourth-order CEF parameters
($B_2^0$, $B_4^0$, and $B_4^4$) are included.
Because of the symmetry reason, the effect of sixth-order terms
($B_6^0$ and $B_6^2$) are dropped.
Thus, $\Gamma_{3}^{+}$ doublet does not appear as a stable ground state
in the $j$-$j$ coupling scheme for $O_{\rm h}$ symmetry.
Moreover, we cannot include the effect of $B_6^2$ which is
characteristic of filled skutterudites with $T_{\rm h}$ symmetry.
\cite{Takegahara}
These points should be improved for the further development
of the microscopic theory of $f$-electron systems
on the basis of the $j$-$j$ coupling scheme.

In this paper, we propose an effective model on the basis of
the $j$-$j$ coupling scheme,
which describes low-energy $f$-electron states
by considering the effect of CEF potentials within the first order
of $1/\lambda$, where $\lambda$ is the spin-orbit interaction.
Since the effect of sixth-order CEF terms ($B_6^0$ and $B_6^2$) are
included as two-body potentials for $f$ electrons,
we can reproduce the CEF energy levels
in the intermediate coupling region on the basis of
the concept of the $j$-$j$ coupling scheme.

The organization of this paper is as follows.
In Sec.~2, first we show the local $f$-electron Hamiltonian $H$
composed of the spin-orbit coupling, the CEF potential,
and the Coulomb interactions.
On the basis of $H$, we discuss the $f$-electron states in
the $LS$ and $j$-$j$ coupling schemes. 
Then, we derive the effective model from $H$
by the perturbation expansion in terms of $1/\lambda$.
In order to compare the results of the effective model
with those of the weak CEF region,
we also define the Stevens Hamiltonian
by the method of operator equivalents.\cite{Stevens}
In Sec.~3, we show the results of the effective model for the cases
of $n$=2$\sim$5 in comparison with those of the Stevens Hamiltonian,
where $n$ denotes the local $f$-electron number.
For each $n$, we compare the CEF energy levels of $H_{\rm eff}$
with those of $H$ in the intermediate coupling region.
We also show that the CEF energy levels for $T_{\rm h}$ symmetry
are reproduced in our modified $j$-$j$ coupling scheme.
In Sec.~4, we briefly discuss the $f$-electron state for $n$$>$7
in the $j$-$j$ coupling scheme,
by showing the CEF energy levels for the cases of $n$=12 and 13.
We remark a couple of future issues concerning the application
of our modified $j$-$j$ coupling schemes.
Finally, the paper is summarized.
Throughout this paper, we use such units as $\hbar$=$k_{\rm B}$=1
and the energy unit is eV.

%
%
\section{Formulation}

\subsection{Hamiltonian}

In general, the local $f$-electron Hamiltonian is given by
\begin{eqnarray}
   H = H_{\rm so} + H_{\rm CEF} + H_{\rm int}.
\end{eqnarray}
The first term indicates the spin-orbit coupling, written as
\begin{eqnarray}
   H_{\rm so} = \lambda \sum_{m,\sigma,m',\sigma'}
   \zeta_{m,\sigma;m',\sigma'} f_{m\sigma}^{\dag}f_{m'\sigma'},
\end{eqnarray}
where $\lambda$ is the spin-orbit interaction,
$f_{m\sigma}$ is the annihilation operator of $f$ electron,
$\sigma$=$+1$ ($-1$) for up (down) spin,
$m$ is the $z$-component of angular momentum $\ell$=3,
and the matrix elements are expressed by
\begin{eqnarray}
 \begin{array}{l}
   \zeta_{m,\sigma;m,\sigma}=m\sigma/2,\\
   \zeta_{m+\sigma,-\sigma;m,\sigma}=\sqrt{\ell(\ell+1)-m(m+\sigma)}/2,
 \end{array}
\end{eqnarray}
and zero for other cases.

The second term denotes the CEF potential, given by
\begin{eqnarray}
  \label{Eq:CEF}
  H_{\rm CEF} = \sum_{m,m',\sigma} B_{m,m'}
  f_{m\sigma}^{\dag}f_{m'\sigma},
\end{eqnarray}
where $B_{m,m'}$ is determined from the CEF table
for $J$=$\ell$=3.\cite{Hutchings}
Note that electrostatic CEF potentials do $not$
act on $f$-electron spin.
For the cubic system with $O_{\rm h}$ symmetry,
$B_{m,m'}$ is expressed by two CEF parameters for $J$=3,
$B_4^0$ and $B_6^0$, as
\begin{eqnarray}
  \begin{array}{l}
    B_{3,3}=B_{-3,-3}=180B_4^0+180B_6^0, \\
    B_{2,2}=B_{-2,-2}=-420B_4^0-1080B_6^0, \\
    B_{1,1}=B_{-1,-1}=60B_4^0+2700B_6^0, \\
    B_{0,0}=360B_4^0-3600B_6^0, \\
    B_{3,-1}=B_{-3,1}=60\sqrt{15}(B_4^0-21B_6^0),\\
    B_{2,-2}=300B_4^0+7560B_6^0.
  \end{array}
\end{eqnarray}
Note the relation of $B_{m,m'}$=$B_{m',m}$.
Following the traditional notation,\cite{LLW} we define
\begin{eqnarray}
  \label{CEFparam}
  B_4^0=Wx/F(4),~B_6^0=W(1-|x|)/F(6),
\end{eqnarray}
where $x$ and the sign of $W$ specify the CEF energy scheme,
while $|W|$ determines the energy scale for the CEF potential.
Concerning non-dimensional parameters, $F(4)$ and $F(6)$,
we choose $F(4)$=15 and $F(6)$=180 for $J$=3
by following Ref.~\citen{LLW}
In actual $f$-electron materials, the magnitude of the CEF potential
is considered to be $10^{-4}$$\sim$$10^{-3}$eV.
For the calculation of the CEF energy level in this paper,
we set $W$=$-10^{-4}$eV,
even if we do not explicitly mention the value of $W$.

Finally, $H_{\rm int}$ denotes Coulomb interaction term, given by
\begin{equation}
   H_{\rm int} \!=\! \sum_{m_1\sim m_4} \! \sum_{\sigma,\sigma'}
   I_{m_1m_2,m_3m_4}
   f_{m_1\sigma}^{\dag}f_{m_2\sigma'}^{\dag}
   f_{m_3\sigma'}f_{m_4\sigma},
\end{equation}
where the Coulomb integral $I_{m_1m_2,m_3m_4}$ is expressed by
\begin{eqnarray}
  I_{m_1m_2,m_3m_4} = \sum_{k=0}^{6} F^k c_k(m_1,m_4)c_k(m_2,m_3).
\end{eqnarray}
Here $F^k$ is the Slater-Condon parameter \cite{Slater1,Condon}
and $c_k$ is the Gaunt coefficient \cite{Gaunt,Racah}
which is tabulated in the standard textbooks of quantum
mechanics.\cite{Slater2}
Note that the sum is limited by the Wigner-Eckart theorem to
$k$=0, 2, 4, and 6.

In principle, the Slater-Condon parameters and the spin-orbit interaction
are determined so as to reproduce the spectra of rare-earth and actinide
ions for each value of $n$, but in any case, the Slater-Condon parameters
are considered to be distributed between 1 eV and 10 eV,
while $\lambda$ is in the order of 0.1 eV.
In this paper, in order to discuss conveniently the competition between
the Coulomb interactions and the spin-orbit coupling,
we parameterize $F^k$ as
\begin{eqnarray}
  \label{SCparam}
  F^0=10U,~F^2=5U,~F^4=3U, F^6=U,
\end{eqnarray}
where $U$ is introduced as a scaling parameter.
As long as we set the Slater-Condon parameters between 1 eV and 10 eV,
the results obtained in the paper do not change qualitatively,
even if we use different values for $F^k$.
As for the meaning of $U$, roughly speaking, it indicates the order
of the Hund's rule interaction $J_{\rm H}$ among $f$ orbitals,
which is considered to be in the order of a few eV.
However, for the purpose to explain the change between the $LS$ and
$j$-$j$ coupling schemes, we treat $U$ as a free parameter
in the following arguments.

\subsection{$LS$ versus $j$-$j$ coupling schemes}

In order to describe the low-energy $f^n$-electron states of $H$,
there are two typical approaches,
the $LS$ coupling and $j$-$j$ coupling schemes.
In the $LS$ coupling scheme,
first the total spin ${\mib S}$ and total angular momentum ${\mib L}$
are formed by following Hund's rules.
After forming ${\mib S}$ and ${\mib L}$, we include the effect of
spin-orbit interaction $\lambda_{LS}$${\mib L}$$\cdot$${\mib S}$,
where $\lambda_{LS}$=$\lambda/n$ for $n$$<$7
and $\lambda_{LS}$=$-\lambda/(14-n)$ for $n$$>$7.
The ground state multiplet is specified by the total
angular momentum ${\mib J}$, given by ${\mib J}$=${\mib L}$+${\mib S}$.
From simple algebra, the ground-state multiplet is
characterized by $J$=$|L$$-$$S|$ for $n$$<$7,
while $J$=$L$+$S$ for $n$$>$7.

On the other hand, in the $j$-$j$ coupling scheme,
first we include the spin-orbit coupling so as
to define the one-electron state
labelled by the total angular momentum ${\mib j}$,
given by ${\mib j}$=${\mib s}$+${\mib \ell}$,
where ${\mib s}$ denotes one-electron spin.
For $f$ orbitals with $\ell$=3, we immediately obtain an octet with
$j$=7/2(=3+1/2) and a sextet with $j$=5/2(=3$-$1/2),
which are well separated by the spin-orbit interaction.
Note here that the level for the octet is higher than that of the sextet.
Then, we take into account the effect of Coulomb interactions
to accommodate $n$ electrons among the sextet and/or octet,
leading to the ground-state level in the $j$-$j$ coupling scheme.
For $n$=1$\sim$6, we accommodate $f$ electrons in the $j$=5/2 sextet
to construct the multiplet specified by $J$,
while for $n$=7$\sim$13, $f$ electrons are accommodated
in the $j$=7/2 octet.

As intuitively understood from the above brief explanation,
the $LS$ and $j$-$j$ coupling schemes become exact
in the limit of $U$=$\infty$ and $\lambda$=$\infty$, respectively.
In actual $f$-electron materials, as mentioned above,
$U$($\approx$$J_{\rm H}$) is as large as a few eV.
On the other hand, for rare-earth atoms,
$\lambda$ is about 0.1$\sim$0.2 eV,
while it is 0.2$\sim$0.3 eV for actinide atoms.
Thus, the actual situation is characterized by $\lambda/U$
in the order of 0.1.
This intermediate coupling region is close to
neither the $LS$ coupling limit
($U$=$\infty$) and the $j$-$j$ coupling limit ($\lambda$=$\infty$),
but due to the relation of $U$$>$$\lambda$ in actual materials,
the $LS$ coupling scheme was widely used for the description
of the local $f$-electron state.
Contrary to such a historical trend,
we believe that it is meaningful to reexamine
the significance of the $j$-$j$ coupling scheme
and to provide an alternative systematic framework
to understand the CEF state of $f^n$-electron compounds.

\begin{figure}[t]
\begin{center}
\includegraphics[width=8.5truecm]{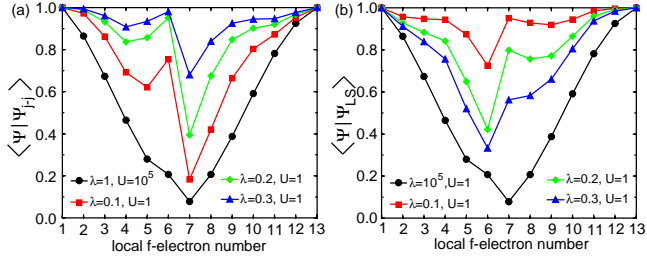}
\caption{Overlap integrals (a) $\langle \Psi | \Psi_{j-j} \rangle$
and (b) $\langle \Psi | \Psi_{LS} \rangle$
for $n$=1$\sim$13 for several parameter sets of $U$ and $\lambda$.}
\end{center}
\end{figure}

For the purpose, let us see how the wavefunction of $H$ is changed
by $U$ and $\lambda$.
Here we suppress the CEF potential to focus on the competition
between the Coulomb interaction and the spin-orbit coupling.
In Fig.~1(a), we depict the overlap integral
$\langle \Psi | \Psi_{j-j} \rangle$ for $n$=1$\sim$13
for several sets of $\lambda$ and $U$,
where $|\Psi \rangle$ is the ground state of $H_{\rm so}+H_{\rm int}$
and $|\Psi_{j-j} \rangle$ denotes
the state at the limit of $\lambda$=$\infty$.
Here we use $\lambda$=$10^5$ and $U$=1 to express virtually
the $j$-$j$ coupling limit.
The solid circles denote the overlap integral between the
$LS$ and $j$-$j$ coupling schemes,
$\langle \Psi_{LS} | \Psi_{j-j} \rangle$,
where $|\Psi_{LS} \rangle$ denotes the state at the $LS$ coupling limit.
Note that we set $\lambda$=1 and $U$=$10^5$, where $U$ is large enough
to arrive at the $LS$ coupling limit.
We find that $\langle \Psi_{LS} | \Psi_{j-j} \rangle$ exhibits
a ``V''-shaped function of $n$.
It is natural, since the Hund's rule interaction works most effectively
at half filling ($n$=7), which is the worst case for
the $j$-$j$ coupling scheme.

In the intermediate coupling region for $\lambda$=0.1$\sim$0.3 and $U$=1,
the overlap integrals are increased with the increase of $\lambda$.
In particular, for $n$$<$7 and $\lambda$=0.1, the values of
$\langle \Psi | \Psi_{j-j} \rangle$ are larger than 0.6.
This lower limit becomes 0.8 and 0.9 for $\lambda$=0.2 and 0.3,
respectively.
These facts suggest that the $j$-$j$ coupling scheme is
the good starting approximation to consider the $f^n$-electron
state for $n$$<$7.

Note that the values of $\langle \Psi | \Psi_{j-j} \rangle$
for $n$=5 and 6 become larger than that for $n$=4,
leading to a shallow minimum around at $n$=3$\sim$4,
when $\lambda$ is increased.
It is intuitively understood by the electron-hole relation
among the $j$=5/2 sextet in the $j$-$j$ coupling scheme.
For $n$=6, the singlet ($J$=0) of fully occupied $j$=5/2 state
is considered to be very stable, since it is the closed-shell structure
in the $j$-$j$ coupling scheme.
For $n$=5, when we accommodate five $f$ electrons in the $j$=5/2 sextet,
it corresponds to the one-hole case.
In the $j$-$j$ coupling limit, we expect that the wavefunction for
one-hole state is the same as that of one-electron case due to the
electron-hole conversion.
Thus, the overlap integral is rather increased for $n$=5 and 6,
in comparison with the case of $n$=4, when we increase $\lambda$.
This fact also supports that we can construct the ground state
in the intermediate coupling region on the basis of
the $j$-$j$ coupling scheme.
Note that for $n$$\ge$7, the overlap integral is also increased
with the increase of $\lambda$ and $n$.
In the region of $n$$\ge$8, the values seem to approach unity
gradually, but for $n$=7, it is increased only slowly.
For the half-filling case, it is difficult to recommend
the usage of the $j$-$j$ coupling scheme.

In Fig.~1(b), we show the overlap integral with the $LS$ coupling
wavefunction, defined by $\langle \Psi | \Psi_{LS} \rangle$,
for $n$=1$\sim$13 for several sets of $\lambda$ and $U$.
The solid circles are the same as those in Fig.~1(a),
since they denote $\langle \Psi_{j-j} | \Psi_{LS} \rangle$.
For $\lambda$=0.1 and $U$=1, the ground state of $H$ is well
approximated by the $LS$ coupling state except for $n$=6,
since we find $\langle \Psi | \Psi_{j-j} \rangle$$>$0.9
for $n$=1$\sim$5 and 7$\sim$13.
When we increase $\lambda$, the overlap integrals are totally
suppressed.
For $\lambda$=0.3, it seems to be difficult to use the $LS$ coupling
scheme, in particular, for the region near $n$=6.

In order to construct an effective model
in the intermediate coupling region
with $\lambda/U$ in the order of 0.1,
it is necessary to improve one of the
$LS$ and $j$-$j$ coupling schemes.
We prefer one of the schemes,
depending on the nature of the current problem.
For $n$$\le$6 and $n$$\ge$8, we believe that
the $j$-$j$ coupling scheme provides us the good starting
point to approach the intermediate coupling region.
As mentioned in the introduction, it is difficult
to include the itinerant nature of $f$ electrons
in the $LS$ coupling scheme.
In order to describe the Bloch state of $f$ electron,
we should prefer the $j$-$j$ coupling scheme
for the description of the $f$-electron state.
Then, we consider the local $f$-electron problem to construct the
effective model by the perturbation expansion in terms of $1/\lambda$
from the limit of $\lambda$=$\infty$.

In the $j$-$j$ coupling scheme, even if we include the effect of the
order of $1/\lambda$, the local $f$-electron state is still composed
of the $j$=5/2 sextet.
When the periodic system is discussed by using the effective model
constructed from the $j$=5/2 states, we consider
the hybridization between $j$=5/2 states and conduction bands,
although in actuality, there should also exist
the hybridization of $j$=7/2 states.
However, in the relativistic band-structure calculations
for light rare-earth materials,
the $j$=5/2 sextet is occupied, while the $j$=7/2 octet is unoccupied,
suggesting that $f$ electrons in the $j$=5/2 sextet mainly contribute
to the ground and the low-energy excited states.
In fact, the Fermi-surface sheets are composed of $j$=5/2 electrons.
Since we are interested in the electronic structure near the Fermi level,
it seems natural to exploit the $j$=5/2 states for the construction
of the effective model to discuss low-energy electronic properties
of $f$-electron materials.

\subsection{Effective model due to the expansion in terms of $1/\lambda$}

Now let us construct the effective Hamiltonian
in the $j$-$j$ coupling scheme
by including the effect of the CEF potentials
in the perturbation expansion in terms of $1/\lambda$.
Note that the energy scale $|W|$ of the CEF potential
is much smaller than $U$ and $\lambda$.
As already mentioned, we set $W$=$-10^{-4}$ eV in this paper
to be consistent with such a situation of the weak CEF.

First we transform the $f$-electron basis
between $(m,\sigma)$ and $(j,\mu)$ representations,
connected by Clebsch-Gordan coefficients,
where $j$ is the total angular momentum and
$\mu$ is the $z$-component of $j$.
Hereafter we use symbols ``$a$'' and ``$b$'' for $j$=5/2 and 7/2,
respectively.
When we define $f_{j\mu}$ as the annihilation operator
for $f$ electron labeled by $j$ and $\mu$,
the transformation is defined as
\begin{eqnarray}
  f_{j\mu} = \sum_{m,\sigma} C_{j,\mu;m,\sigma} f_{m\sigma},
\end{eqnarray}
where the Clebsch-Gordan coefficient $C_{j,\mu;m,\sigma}$ is give by
\begin{eqnarray}
 \begin{array}{l}
  C_{a,\mu;\mu-\sigma/2,\sigma}=-\sigma \sqrt{(7/2-\sigma \mu)/7},\\
  C_{b,\mu;\mu-\sigma/2,\sigma}=\sqrt{(7/2+\sigma \mu)/7},
 \end{array}
\end{eqnarray}
and other components are zero.

After the transformation, we express $H$ as
\begin{equation}
  H = H_{\rm so}+H',
\end{equation}
where the spin-orbit coupling term $H_{\rm so}$ is diagonalized as
\begin{equation}
  H_{\rm so} = \sum_{j,\mu}
  \lambda_{j}f^{\dag}_{j\mu}f_{j\mu},
\end{equation}
with $\lambda_a$=$-2\lambda$ and $\lambda_b$=$(3/2)\lambda$.
The remaining part includes the CEF and Coulomb interaction terms as
\begin{equation}
  H'=H_{\rm CEF}+H_{\rm int},
\end{equation}
where $H_{\rm CEF}$ and $H_{\rm int}$ are given by
\begin{equation}
  H_{\rm CEF} = \sum_{j_1\mu_1,j_2\mu_2}
  {\tilde B}^{j_1,j_2}_{\mu_1,\mu_2}
  f^{\dag}_{j_1\mu_1}f_{j_2\mu_2},
\end{equation}
and
\begin{equation}
  H_{\rm int}=\sum_{j_1 \sim j_4}\sum_{\mu_1 \sim \mu_4}
  {\tilde I}^{j_1j_2,j_3j_4}_{\mu_1\mu_2,\mu_3\mu_4}
  f^{\dag}_{j_1\mu_1}f^{\dag}_{j_2\mu_2}f_{j_3\mu_3}f_{j_4\mu_4},
\end{equation}
respectively.
Note that ${\tilde B}$ and ${\tilde I}$ are the CEF potential
and Coulomb interactions expressed in the basis of $j$ and $\mu$.

In order to obtain the effective model,
we use the degenerate perturbation theory
by treating $H'$ as a perturbation to $H_{\rm so}$.
The effective model is written as
\begin{equation}
  H_{\rm eff}=H_a^{(0)}+H_a^{(1)},
\end{equation}
where $H^{(k)}_j$ denotes the $k$th-order term
with respect to $1/\lambda$ for the multiplet labeled by $j$.
The zeroth-order term $H^{(0)}_a$ indicates the model
in the standard $j$-$j$ coupling scheme.\cite{Hotta1a,Hotta1b}
When we simply ignore all the terms including the symbol $b$ ($j$=7/2),
we obtain $H^{(0)}_a$ as
\begin{equation}
  H^{(0)}_a = H_{\rm CEF}^{(0)} + H_{\rm int}^{(0)}.
\end{equation}
The CEF term is given by
\begin{equation}
  H_{\rm CEF}^{(0)} = \sum_{\alpha_1,\alpha_2}
    {\tilde B}^{a,a}_{\alpha_1,\alpha_2}
    f^{\dag}_{a\alpha_1}f_{a\alpha_2},
\end{equation}
where ${\tilde B}^{a,a}_{\alpha_1,\alpha_2}$ denotes
the CEF potential for $J$=5/2, given by
\begin{eqnarray}
  \label{CEFjj}
  \begin{array}{l}
    {\tilde B}^{a,a}_{\pm 5/2,\pm 5/2} =  60 B_4^0(1,5/2), \\
    {\tilde B}^{a,a}_{\pm 3/2,\pm 3/2} = -180 B_4^0(1,5/2), \\
    {\tilde B}^{a,a}_{\pm 1/2,\pm 1/2} =  120 B_4^0(1,5/2), \\
    {\tilde B}^{a,a}_{\pm 5/2,\mp 3/2} = B_{\mp 3/2,\pm 5/2}=
    60 \sqrt{5} B_4^0(1,5/2).
  \end{array}
\end{eqnarray}
Here $B_4^0(n,J)$ denotes the CEF parameter for $n$ and $J$,
where $n$ is the local $f$-electron number and $J$ denotes
the total angular momentum of ground state multiplet.
For $n$=1 and $J$=5/2, $B_4^0(1,5/2)$ is related with $B_4^0$
for $J$=$\ell$=3 in eq.~(\ref{CEFparam}) as
\begin{equation}
   \label{CEFjj1}
   B_4^0(1,5/2)=(11/7)B_4^0.
\end{equation}
The meaning of the coefficient $11/7$ will be explained later.

The Coulomb interaction term is given by
\begin{equation}
  H_{\rm int}^{(0)}=\sum_{\alpha_1 \sim \alpha_4}
  {\tilde I}^{aa,aa}_{\alpha_1\alpha_2,\alpha_3\alpha_4}
  f^{\dag}_{a\alpha_1}f^{\dag}_{a\alpha_2}f_{a\alpha_3}f_{a\alpha_4},
\end{equation}
where the Coulomb integral ${\tilde I}^{aa,aa}$ is expressed by
three Racah parameters, $E_0$, $E_1$, and $E_2$,\cite{Hotta1a,Hotta1b}
which are related to the Slater-Condon parameters as
\begin{eqnarray}
  \label{Racahjj}
  \begin{array}{l}
    E_0 = F^0-(80/1225)F^2-(12/441)F^4,\\
    E_1 = (120/1225)F^2+(18/441)F^4, \\
    E_2 = (12/1225)F^2-(1/441)F^4.
  \end{array}
\end{eqnarray}
Among Coulomb interactions, the Hund's rule coupling is
the relevant energy scale for the construction of multi-$f$-state.
In this case, among three Racah parameters, $E_2$ plays
a role of the Hund's rule coupling in the $j$-$j$ coupling scheme.
Concerning the magnitude,
by combining eqs.~(\ref{SCparam}) and (\ref{Racahjj}),
we obtain $E_2$=(31/15)$U$/49=$0.04U$
in the present parameterization.
Note that the magnitude of the Hund's rule interaction
in the $j$-$j$ coupling scheme is generally reduced.
In order to understand this point intuitively,
it is convenient to consider a simple Hund's rule term,
expressed as $-J_{\rm H} {\mib s}^2$,
where ${\mib s}$ denotes the spin of $f$ electron.
Here we note the relation ${\mib s}$=$(g_J-1){\mib j}$,
where $g_J$ is the Land\'e's $g$-factor and ${\mib j}$
indicates the total angular momentum.
Since $g_J$=6/7 for $j$=5/2, we obtain ${\mib s}$=$-(1/7){\mib j}$.
Namely, the original Hund's rule term is rewritten as
$-(J_{\rm H}/49) {\mib j}^2$ in the $j$-$j$ coupling scheme.
Thus, the Hund's rule interaction in the $j$-$j$ coupling scheme
is reduced as $J_{\rm H}/49$ from the original value.
As discussed later, this reduction has an important meaning
to understand the rather wide applicability of the effective model.

Now we consider the correction term in the order of $1/\lambda$,
which is written as
\begin{equation}
  H_a^{(1)}=H^{(1)}_{\rm CEF}+H^{(1)}_{\rm int}.
\end{equation}
The CEF term is given by
\begin{equation}
  H^{(1)}_{\rm CEF} = \sum_{\alpha_1,\alpha_2}
  {\tilde B}^{(1)}_{\alpha_1,\alpha_2}
  f^{\dag}_{a\alpha_1}f_{a\alpha_2},
\end{equation}
with
\begin{eqnarray}
  {\tilde B}^{(1)}_{\alpha_1,\alpha_2} = \sum_{\beta}
  \frac{\langle \alpha_1 | H_{\rm CEF} |\beta \rangle
  \langle \beta | H_{\rm CEF} |\alpha_2 \rangle}
  {\lambda_a-\lambda_b},
\end{eqnarray}
where $|\alpha \rangle$=$f^{\dag}_{a\alpha}|0\rangle$ and
$|\beta \rangle$=$f^{\dag}_{b\beta}|0\rangle$.
For $O_{\rm h}$ symmetry, after some algebraic calculations,
we obtain
\begin{eqnarray}
  \begin{array}{l}
    {\tilde B}^{(1)}_{5/2,5/2} ={\tilde B}^{(1)}_{-5/2,-5/2}
    =5 \varepsilon_8/6 + \varepsilon_7/6, \\
    {\tilde B}^{(1)}_{3/2,3/2} ={\tilde B}^{(1)}_{-3/2,-3/2}
    =\varepsilon_8/6 + 5 \varepsilon_7/6, \\
    {\tilde B}^{(1)}_{1/2,1/2} ={\tilde B}^{(1)}_{-1/2,-1/2}
    =\varepsilon_8, \\
    {\tilde B}^{(1)}_{5/2,-3/2} ={\tilde B}^{(1)}_{3/2,-5/2}
    =\sqrt{5}(\varepsilon_8-\varepsilon_7)/6,
  \end{array}
\end{eqnarray}
where $\varepsilon_7$ and $\varepsilon_8$ are,
respectively, given by
\begin{equation}
 \varepsilon_7=-\Bigl(\frac{240}{7}\Bigr)^2 \frac{6}{7\lambda}
  (5B_4^0+126 B_6^0)^2,
\end{equation}
and
\begin{equation}
 \varepsilon_8=-\Bigl(\frac{720}{7}\Bigr)^2 \frac{10}{7\lambda}
  (B_4^0-21 B_6^0)^2.
\end{equation}
These values indicate the energy corrections for $\Gamma_7^-$ and
$\Gamma_8^-$ states at $n$=1, respectively.
Note that such corrections are in the order of $W^2/\lambda$,
as easily understood from the definition of
${\tilde B}^{(1)}_{\alpha_1,\alpha_2}$.

Concerning the $1/\lambda$-correction to the two-body potential,
$H^{(1)}_{\rm int}$ is expressed as
\begin{equation}
   H^{(1)}_{\rm int} = \sum_{\alpha_1 \sim \alpha_4}
   {\tilde I}^{(1)}_{\alpha_1\alpha_2,\alpha_3\alpha_4}
   f^{\dag}_{a\alpha_1}f^{\dag}_{a\alpha_2}
   f_{a\alpha_3}f_{a\alpha_4},
\end{equation}
where the two-body potential is formally given by
\begin{equation}
   \label{eq:int1}
   \begin{split}
   {\tilde I}^{(1)}_{\alpha_1\alpha_2,\alpha_3\alpha_4} & =
   \sum_{\alpha,\beta}
   \frac{\langle \alpha_1 \alpha_2 | H' | \alpha \beta \rangle
   \langle \beta \alpha | H' |\alpha_3 \alpha_4 \rangle}
   {\lambda_a-\lambda_b} \\
   &+ \sum_{\beta_1,\beta_2}
   \frac{\langle \alpha_1 \alpha_2 | H' |\beta_1\beta_2\rangle
   \langle \beta_2 \beta_1| H' |\alpha_3 \alpha_4 \rangle}
   {2(\lambda_a-\lambda_b)} \\
   &-\langle \alpha_1 \alpha_2 | H^{(1)}_{\rm CEF} | \alpha_3\alpha_4 \rangle,
  \end{split}
\end{equation}
where
$|\alpha_1 \alpha_2 \rangle$=
$f^{\dag}_{a\alpha_1}f^{\dag}_{a\alpha_2}|0\rangle$,
$|\alpha \beta \rangle$=
$f^{\dag}_{a\alpha}f^{\dag}_{b\beta}|0\rangle$,
and
$|\beta_1\beta_2 \rangle$=
$f^{\dag}_{b\beta_1}f^{\dag}_{b\beta_2}|0\rangle$.
Note that the final term in eq.~(\ref{eq:int1}) is needed
to avoid the double-count of the contributions
from $H^{(1)}_{\rm CEF}$,
when we diagonalize $H_{\rm eff}$ for $n$$\ge$2.
As we will see later, the two-body effective potential
${\tilde I}^{(1)}$ plays a crucial role
in the intermediate coupling region.
It may be possible to obtain the analytic form of ${\tilde I}^{(1)}$
after lengthy and tedious algebraic calculations,
but in this paper, we evaluate ${\tilde I}^{(1)}$ only numerically,
since our purpose here is to show that $H_{\rm eff}$ actually works.
The derivation of more convenient analytic form of ${\tilde I}^{(1)}$
is one of future tasks.

Now we discuss the range of the value of $\lambda$,
in which $H_{\rm eff}$ works.
First we note that $H^{(1)}_{\rm CEF}$ is in the order of $W^2/\lambda$.
Since this term is small in the order of $|W|/\lambda$
in comparison with $H^{(0)}_{\rm CEF}$,
it does not play important roles.
The first two terms in eq.~(\ref{eq:int1}) also include
the contributions in the order of $W^2/\lambda$,
but we find that they are exactly cancelled by the last term.
Thus, the lowest contribution to the CEF potential from
$H^{(1)}_{\rm int}$ is in the order of $|W|U/\lambda$.
Note that the terms in the order of $U^2/\lambda$ are not important,
since they contribute to the energy shift of the ground-state multiplet.

From the mathematical viewpoint of the convergence of the perturbation
expansion, $|W|U/\lambda$ is thought to be smaller than $|W|$,
which is the energy scale of $H^{(0)}_{\rm CEF}$.
Namely, we obtain the condition of $U/\lambda$$\ll$1,
but under this condition, $H_{\rm eff}$ cannot be used
for realistic systems at the first glance,
since $U$ is larger than $\lambda$ in $f$-electron compounds.
In order to reconsider this point,
we note that the sixth-order contributions of the CEF potential
first appear in $H^{(1)}_{\rm CEF}$ in the order of $|W|U/\lambda$.
Since it is enough for us to keep the condition of the weak CEF,
$|W|U/\lambda$ should be smaller than the relevant energy scale
of $H^{(0)}_{\rm int}$, i.e., the Hund's rule interaction
in the $j$-$j$ coupling scheme.
Thus, we obtain another realistic condition of
$|W|U/\lambda$$\ll$$E_2$.
Here we note that the right-hand side is $E_2$, not $U$.
On the other hand, in the left-hand side, $U$ is needed,
since all the Coulomb interaction terms contribute
to eq.~(\ref{eq:int1}) in the intermediate processes.
From eqs.~(\ref{SCparam}) and (\ref{Racahjj}),
we obtain the revised condition of $|W|/\lambda$$\ll$0.04
in the present parameter choice.
In comparison with one of the original conditions of
the weak CEF ($|W|/\lambda$$\ll$1), the range of $\lambda$
in which $H_{\rm eff}$ works seems to be narrow,
but when we compare the condition of $|W|/\lambda$$\ll$0.04
with that of $U/\lambda$$\ll$1,
we understand that the range of $\lambda$ becomes wide
so as to include the realistic parameter space.
For instance, for the case of $|W|$=$10^{-4}$eV,
it is allowed to take $\lambda$ in the order of 0.1eV,
which is the realistic value for rare-earth ion.

\subsection{Stevens Hamiltonian in the weak CEF}

In the previous subsection, we have set the effective model $H_{\rm eff}$
due to the expansion in terms of $1/\lambda$ from the limit of
$\lambda$=$\infty$.
In order to assess the applicability of the effective model
from the quantitative viewpoint,
it is necessary to compare the results of $H_{\rm eff}$
with those of $H$.
For the purpose, in addition to the direct comparison between
$H$ and $H_{\rm eff}$, it is useful to introduce the CEF Hamiltonian
for $O_{\rm h}$ symmetry in the region of weak CEF, since we always
consider the case of $U$$\gg$$|W|$ and $\lambda$$\gg$$|W|$.
The CEF Hamiltonian is conventionally expressed by using
the method of Stevens' operator equivalent as~\cite{Stevens}
\begin{equation}
  H_{\rm S} \!=\! B_4^0(n,J)({\hat O}_4^0+5{\hat O}_4^4)
  \!+\! B_6^0(n,J)({\hat O}_6^0-21{\hat O}_6^4),
\end{equation}
where $B_p^q(n,J)$ and ${\hat O}_p^q$ denote, respectively,
the CEF parameter and the Stevens' operator equivalent for $n$ and $J$.
We call $H_{\rm S}$ the Stevens Hamiltonian.
The matrix elements of ${\hat O}_p^q$ for any value of $J$ have
been tabulated by Hutchings.\cite{Hutchings}
Here we explicitly show the values of $n$ and $J$ in the parentheses
of the CEF parameter, in order to distinguish them from
$B_4^0$ and $B_6^0$ for $J$=$\ell$=3 in eq.~(\ref{CEFparam}).
Note that $H_{\rm S}$ is the effective Hamiltonian for the multiplet
specified by $J$ for any values of $U$ and $\lambda$,
as long as they are much larger than $|W|$.
In fact, we have checked that the energy levels of $H$ are correctly
reproduced by $H_{\rm S}$ with satisfactory precision for $U$($\gg$$|W|$)
and $\lambda$($\gg$$|W|$),
except for the case of $n$=3 and $\lambda$=$\infty$.

Effects of $U$ and $\lambda$ appear in the CEF parameters,
$B_4^0(n,J)$ and $B_6^0(n,J)$, which are expressed by
\begin{equation}
   \label{eq:CEF2}
   B_4^0(n,J) = A_4 \langle r^4 \rangle \beta^{(n)}_{J},~
   B_6^0(n,J) = A_6 \langle r^6 \rangle \gamma^{(n)}_{J},
\end{equation}
where $A_k$ is the parameter depending on materials,
$\beta^{(n)}_{J}$ and $\gamma^{(n)}_{J}$ are the so-called
Stevens factors, which are coefficients
appearing in the method of Stevens' operator equivalent,\cite{Stevens}
and $\langle r^k \rangle$ denotes the radial average of
local $f$-electron wavefunction.
Note that in general, $\beta^{(n)}_{J}$ and $\gamma^{(n)}_{J}$
depend on the Coulomb interaction and spin-orbit coupling,
since they are determined by the nature of the ground-state multiplet
specified by $J$.
In the present paper, the CEF potentials are always given by
$B_4^0$ and $B_6^0$.
Thus, we express $B_4^0(n,J)$ and $B_6^0(n,J)$ as
\begin{equation}
   \label{eq:CEF1}
   B_4^0(n,J) =k_{4}(n,J)B_4^0,~
   B_6^0(n,J) =k_{6}(n,J)B_6^0.
\end{equation}
By assuming that $A_k$ and $\langle r^k \rangle$ are
not changed in the same material group, we obtain
\begin{equation}
   \label{eq:CEF3}
   k_4(n,J)=\beta^{(n)}_{J}/\beta_{\ell},~
   k_6(n,J)=\gamma^{(n)}_{J}/\gamma_{\ell},
\end{equation}
where $\beta_{\ell}$ and $\gamma_{\ell}$ for one $f$ electron
with $\ell$=3 are given by~\cite{Stevens}
\begin{equation}
   \beta_{\ell}=2/(45\cdot11),~
   \gamma_{\ell}=-4/(9\cdot13\cdot33),
\end{equation}
respectively.

First let us consider the limit of $U$=$\infty$,
i.e., the $LS$ coupling scheme.
We easily obtain $k_4(n,J)$ and $k_6(n,J)$ due to simple algebraic
calculations by using the values of $\beta^{(n)}_{J}$ and
$\gamma^{(n)}_{J}$ for the $LS$ coupling scheme.\cite{Stevens}
For $n$=1 and $J$=5/2, from $\beta^{(1)}_{5/2}$=2/(45$\cdot$7),
we easily obtain
\begin{equation}
   k^{LS}_{4}(1,5/2)=11/7,
\end{equation}
as shown in eq.~(\ref{CEFjj1}). For $n$=2 and $J$=4, we find
\begin{eqnarray}
  \label{eq:CEF4}
   \begin{array}{l}
    k^{LS}_4(2,4)=\beta^{(2)}_{4}/\beta_{\ell}=-2/11,\\
    k^{LS}_6(2,4)=\gamma^{(2)}_{4}/\gamma_{\ell}=-68/1155.
   \end{array}
\end{eqnarray}
We can obtain $k^{LS}_4(n,J)$ and $k^{LS}_6(n,J)$ for
$n$$\ge$3, as shown in Table I.

\begin{table}
\begin{tabular}{c|c||c|c||c|c} \hline
 & & \multicolumn{2}{c||}{$k_4(n,J)$} &
\multicolumn{2}{c}{$k_6(n,J)$} \\ \cline{3-6}
\multicolumn{1}{c|}{\raisebox{1.5ex}[0pt]{$n$}} &
\multicolumn{1}{c||}{\raisebox{1.5ex}[0pt]{$J$}} & 
\multicolumn{1}{c|}{$LS$} &
\multicolumn{1}{c||}{$j$-$j$} &
\multicolumn{1}{c|}{$LS$} &
\multicolumn{1}{c}{$j$-$j$} \\
\hline
  $1$ & $5/2$ &   $11/7$    &   $11/7$ & $0$ & $0$ \\
\hline
  $2$ &  $4$  &   $-2/11$   & $-11/49$ & $-68/1155$ & $0$ \\
\hline
  $3$ & $9/2$ & $-340/4719$ & $0$ & $1615/44044$ & $0$ \\
\hline
  $4$ &  $4$  &  $476/4719$ & $11/49$ & $-646/11011$ & $0$ \\
\hline
  $5$ & $5/2$ &   $13/21$   & $-11/7$ & $0$ & $0$ \\ \hline
\end{tabular}
\caption{Coefficients $k_4(n,J)$ and $k_6(n,J)$ for $n$=1$\sim$5
both in the $LS$ and $j$-$j$ coupling schemes.}
\end{table}

Next we consider another limit of $\lambda$=$\infty$, i.e.,
the $j$-$j$ coupling scheme.
As long as we consider the situation of the weak CEF,
it is possible to obtain the CEF energy levels
by using the Stevens Hamiltonian $H_{\rm S}$,
even in the limit of $\lambda$=$\infty$,
provided that $B_4^0(n,J)$ and $B_6^0(n,J)$ are correctly evaluated.
First we remark a couple of issues which can be understood without
calculations:
(i) Since $B_6^0$ does not appear in the $j$-$j$ coupling scheme
due to the symmetry reason, we always obtain $k^{j-j}_6(n,J)$=0
for $n$$<$7 in the $j$-$j$ coupling limit.
(ii) For $n$=1, $k^{j-j}_4(1,5/2)$ is equal to $k^{LS}_4(1,5/2)$.

Concerning $k_4(n,J)$ for $n$=2$\sim$5 in the $j$-$j$ coupling limit,
we evaluate the value in the following procedure:
Let us consider the case of $n$=2 as a typical example.
We note $|2,4,4\rangle$=$f_{a,5/2}^{\dag}f_{a,3/2}^{\dag}|0\rangle$
in the $j$-$j$ coupling scheme,
where $|n,J,J_z\rangle$ generally denotes the eigenstate determined
by $H_{\rm so}$+$H_{\rm int}$,
$J_z$ is the $z$ component of total angular momentum $J$,
and $|0\rangle$ is the vacuum state.
In the $j$-$j$ coupling scheme, $|n,J,J_z\rangle$ for $n$$\le$6
indicates the eigenstate of $H_{\rm int}^{(0)}$.
Note that the bra vector is defined as $\langle n,J,J_z|$.
Then, we evaluate the CEF matrix element as
\begin{equation}
 \langle 2,4,4 |H_{\rm CEF}^{(0)}| 2,4,4\rangle
 \!=\! -120B_4^0(1,5/2) \!=\! -(1320/7)B_4^0,
\end{equation}
from eqs.~(\ref{CEFjj}) and (\ref{CEFjj1}).
Since the same matrix element is given by $840B_4^0(2,4)$
from the Hutchings table,\cite{Hutchings} we obtain
\begin{equation}
  k^{j-j}_4(2,4)=-11/49,
\end{equation}
in the $j$-$j$ coupling limit for $n$=2 and $J$=4.
Also for $n$=3$\sim$5, we can evaluate $k^{j-j}_4(n,J)$
in the similar way and the results are listed in Table I.
It should be noted that $k^{j-j}_4(3,9/2)$=0.
Since $|3,9/2,9/2\rangle$=
$f_{a,5/2}^{\dag}f_{a,3/2}^{\dag}f_{a,1/2}^{\dag}|0\rangle$,
we easily find
$\langle 3,9/2,9/2 |H_{\rm CEF}^{(0)}| 3,9/2,9/2\rangle$=0.
Namely, for $n$=3 and $J$=9/2,
the lowest-order contribution of the CEF potential is in the order of $W^2$,
not in the order of $W$, in the $j$-$j$ coupling scheme.
This point will be discussed later again.

For the intermediate coupling region in which both $U$ and $\lambda$
are finite, we evaluate numerically $k_4(n,J)$ and $k_6(n,J)$
by deriving $H_{\rm S}$ from the original Hamiltonian $H$.
First we diagonalize $H_{\rm so}+H_{\rm int}$ as
\begin{equation}
  (H_{\rm so}+H_{\rm int})|n,J,J_z\rangle = E(n,J)|n,J,J_z\rangle.
\end{equation}
Note that the ground state has $(2J+1)$-fold degeneracy,
when the CEF potential is ignored.
Then, we consider the CEF potential of $O_{\rm h}$ symmetry
in the first order of $W$
and the Hamiltonian $H_{\rm S}$ is expressed in the matrix form as
\begin{equation}
  \label{HS2}
  H_{\rm S}(J_z,J'_z)
  =\langle n,J,J_z | H_{\rm CEF} | n,J,J'_z \rangle.
\end{equation}
Since the matrix elements of $H_{\rm S}$ have been already listed
for each value of $J$ by using $B_4^0(n,J)$ and $B_6^0(n,J)$,
\cite{Hutchings}
we can numerically obtain $B_4^0(n,J)$ and $B_6^0(n,J)$
from eq.~(\ref{HS2}) for any values of $U$ and $\lambda$
for a given value of $n$.

For comparison, we also evaluate $B_4^0(n,J)$ and $B_6^0(n,J)$
from $H_{\rm eff}$.
In this case, the procedure is almost the same as above.
First we diagonalize $H_{\rm eff}$ at $W$=0 as
\begin{equation}
  H_{\rm eff}(W\!=\!0)|n,J,J_z\rangle = E_{\rm eff}(n,J)|n,J,J_z\rangle.
\end{equation}
Then, the Hamiltonian $H_{\rm S}^{\rm eff}$ from $H_{\rm eff}$
for small $W$ is expressed in the matrix form as
\begin{equation}
  \label{HSeff}
  \begin{split}
  H_{\rm S}^{\rm eff}(J_z,J'_z)
  &=\langle n,J,J_z | H_{\rm eff}(W)| n,J,J'_z \rangle \\
  &-E_{\rm eff}(n,J)\delta_{J_z,J_z'}.
  \end{split}
\end{equation}
In order to obtain $B_4^0(n,J)$ and $B_6^0(n,J)$ numerically,
we compare the matrix elements of eq.~(\ref{HSeff}) with the table
of Hutchings.

%
%
\section{Results}

\subsection{$f^2$ states}

First we explain the case of $n$=2 in detail.
Before proceeding to the result of $H_{\rm eff}$ for $n$=2,
let us examine the results of the $LS$ and $j$-$j$ coupling schemes.
Then, readers can understand how the $j$-$j$ coupling scheme will
be improved by $H_{\rm eff}$.
In Figs.~2(a) and 2(b), we show the CEF energy levels obtained by
the direct diagonalization of $H$ for ($U$, $\lambda$)=($10^5$,1)
and (1,$10^5$), which correspond to the $LS$ and the $j$-$j$
coupling limits, respectively.
Note that the origin of the energy is appropriately shifted
for convenience.

\begin{figure}[t]
\begin{center}
\includegraphics[width=8.5truecm]{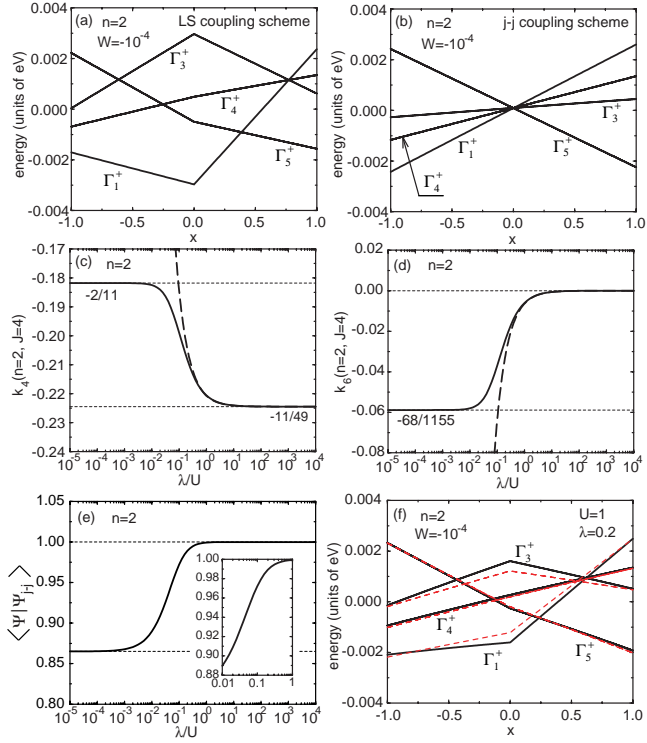}
\caption{(a) CEF energy levels in the $LS$ coupling scheme for $n$=2.
We diagonalize $H$ for $U$=$10^5$ and $\lambda$=1.
(b) CEF energy levels in the $j$-$j$ coupling scheme for $n$=2.
Here we set $U$=1 and $\lambda$=$10^5$ for $H$.
(c) $k_4(2,4)$ and (d) $k_6(2,4)$ as functions of
$\lambda/U$ for $n$=2.
Broken curves denote $k_4(2,4)$ and $k_6(2,4)$
evaluated from $H_{\rm eff}$.
(e) Overlap integral $\langle \Psi | \Psi_{j-j} \rangle$
as a function of $\lambda/U$ for $n$=2.
Inset indicates the result in the intermediate coupling region
in a magnified scale.
The overlap is evaluated by the diagonalization of $H_{\rm so}+H_{\rm int}$.
(f) CEF energy levels of $H_{\rm eff}$ (black solid curves) and
those of $H$ (red broken curves) for $n$=2, $U$=1, and $\lambda$=0.2.}
\end{center}
\end{figure}

In Fig.~2(a), the situation corresponds to the $LS$ coupling scheme,
in which first we obtain the ground-state level
as $^{3}H$ with $S$=1 and $L$=5 due to the Hund's rules in $f$ shells.
Upon further including the spin-orbit interaction,
the ground state is specified by $J$=4 expressed
as $^3H_4$ in the traditional notation.
After considering the CEF potential,
we find nine eigen states, $\Gamma_{1}^{+}$ singlet,
$\Gamma_{3}^{+}$ doublet, and two kinds of triplets
($\Gamma_{4}^{+}$ and $\Gamma_{5}^{+}$).\cite{LLW}
It is found that the eigen energies of $H_{\rm S}$ for $n$=2 and $J$=4
agree perfectly with those in Fig.~2(a), when we use
$k_4^{LS}(2,4)$ and $k_6^{LS}(2,4)$ in Table I.

In Fig.~2(b), we show the results of $H$
in the limit of the $j$-$j$ coupling scheme.
We have checked that these CEF energies are the same as those of 
$H^{(0)}_a$ for $n$=2.
Note that the results seem to be different from those
in Ref.~\citen{Hotta1a} at the first glance,
but it is simply due to the negative sign of $W$.
As mentioned in the previous section,
even for the $j$-$j$ coupling scheme, as long as we consider the
situation of the weak CEF, it is possible to obtain the CEF energy
levels by using the Stevens Hamiltonian $H_{\rm S}$.
We obtain the same results as those in Fig.~2(b)
by diagonalizing $H_{\rm S}$ with $B_4^0(2,4)$=$-(11/49)B_4^0$
and $B_6^0(2,4)$=0.

The energy levels in Fig.~2(b) can be understood more intuitively as follows.
In the limit of $\lambda$=$\infty$, the effective model $H_{\rm eff}$ is
simply reduced to $H^{(0)}_a$, in which the only relevant CEF parameter is
$B_4^0$, leading to the splitting between $\Gamma_{7}^{-}$ and
$\Gamma_{8}^{-}$ levels for $n$=1.
Thus, we obtain the eigen states for $n$=2 by accommodating a couple of
electrons in the potential for $n$=1.
For $x$$<$0 with positive $B_4^0$, we find the $\Gamma_{1}^{+}$ singlet
ground state, including doubly occupied $\Gamma_{7}^{-}$ orbitals,
since $\Gamma_{7}^{-}$ doublet is the ground state for $n$=1
in the region of $x$$<$0.
The first excited state is $\Gamma_{4}^{+}$ triplet, which is formed by
$\Gamma_{7}^{-}$ and $\Gamma_{8}^{-}$ electrons.
For $x$$>$0, since the ground state for $n$=1 is the $\Gamma_{8}^{-}$
quartet, the ground state for $n$=2 is $\Gamma_{5}^{+}$ triplet
composed of a couple of $\Gamma_{8}^{-}$ electrons,
stabilized by the Hund's rule interaction.
The first excited state in this region is $\Gamma_{3}^{+}$ doublet,
which contains the component of a couple of $\Gamma_{8}^{-}$ electrons.

It is true that all the states appearing in the $LS$ coupling scheme
can be also found in the $j$-$j$ coupling scheme.
In particular, at $x$=$\pm$1, the CEF energy levels agree well with
those in Fig.~2(a).
However, when we compare Figs.~2(a) and 2(b) in the region of $|x|$$\ne$1,
the $j$-$j$ coupling results seem to be different from those of
the $LS$ coupling scheme, since the effect of $B_6^0$ is not included,
as already mentioned in the introduction and in Ref.~\citen{Hotta1a}.
This is quite natural, if we recall the fact that there exist
no contributions of $B_6^0$ in the Hutchings table for $J$=5/2.
This point is improved when we consider the $1/\lambda$ corrections,
as shown later.

In Figs.~2(c) and 2(d), we show $k_4(2,4)$ and $k_6(2,4)$, respectively,
as functions of $\lambda/U$.
Since the results have been found to depend only on the ratio of $\lambda/U$,
we change $\lambda$ and $U$ appropriately by keeping the condition
of the weak CEF, i.e., $\lambda$$\gg$$|W|$ and $U$$\gg$$|W|$.
In actual calculations, for $\lambda/U$$\ge$1, we gradually increase
$\lambda$ from unity by setting $U$=1, while for $\lambda/U$$\le$1,
$U$ is increased from unity for the fixed value of $\lambda$=1.
Both in the limits of $\lambda/U$=0 ($LS$ coupling scheme) and
$\lambda/U$=$\infty$ ($j$-$j$ coupling scheme),
the numerical results correctly approach the analytic values
shown in Table I.
For the intermediate coupling region of $\lambda/U$$\sim$0.1,
$k_{4}(2,4)$ and $k_{6}(2,4)$ smoothly change
between the $LS$ and $j$-$j$ coupling values.
We note that $k_4(2,4)$$<$0 and $k_6(2,4)$$<$0,
suggesting that the sign of $B_4^0(2,4)$ and $B_6^0(2,4)$ should be
different from that of $B_4^0$.
In order to set $B_4^0(2,4)$$>$0 and $B_6^0(2,4)$$>$0
for the comparison with the results in Ref.~\citen{LLW},
the sign of $W$ is taken as negative.

Here we remark that $k_{4}(2,4)$ and $k_{6}(2,4)$ become good
measures to show how the situation is close to the $LS$ or
the $j$-$j$ coupling scheme.
For actual materials, it is considered that $U$ is about 1eV,
while $\lambda$ is in the order of 0.1 eV,
leading to $\lambda/U$ in the order of 0.1.
From Figs.~2(c) and 2(d), $k_{4}(2,4)$ and $k_{6}(2,4)$
in such a region are close to neither the value of
the $LS$ nor that of the $j$-$j$ coupling scheme.
Namely, the actual situation corresponds
to the intermediate coupling region.
In particular, we note that the $LS$ coupling scheme is different
from the actual situation, in contrast to the naive expectation
for the validity of the $LS$ coupling scheme.
One may consider that we can fit the results of
the intermediate coupling region by changing $B_4^0(n,J)$
and $B_6^0(n,J)$ in the $LS$ coupling scheme.
It is true that we can change $A_4$ and $A_6$ as fitting parameters,
but it is not allowed to fit the Stevens factors,
which are determined by $U$ and $\lambda$
under the symmetry constraint.
Thus, the actual situation of the intermediate coupling region
is different both from the $LS$ and $j$-$j$ coupling schemes.

However, the above results on the smooth change of $B_4^0(2,4)$ and
$B_6^0(2,4)$ seem to indicate that
the $LS$ and $j$-$j$ coupling schemes are continuously connected.
In fact, when we evaluate the overlap integral concerning
the ground-state wavefunction, the magnitude is continuously changed
between the $LS$ and $j$-$j$ coupling schemes.
In Fig.~2(e), we show the overlap integral
$\langle \Psi | \Psi_{j-j} \rangle$ as a function of $\lambda/U$.
When we decrease $\lambda/U$, we find that the overlap integral is
gradually decreased, but even at the limit of $U$=$\infty$,
we still find $\langle \Psi_{LS} | \Psi_{j-j} \rangle$=0.865.
Namely, the $f^2$-state composed of a couple of electrons in the
$j$=5/2 sextet becomes the good approximation for the $LS$ coupling scheme,
even though the $j$=7/2 octet is simply discarded.

In principle, it is possible to approach the intermediate
coupling region either from the $LS$ or the $j$-$j$ coupling limit,
as mentioned in Sec.~2.2.
Depending on the nature of the problem, we can prefer one of them,
but in this paper, the $j$-$j$ coupling scheme is chosen.
Here we remark that $\langle \Psi | \Psi_{j-j} \rangle$ is close
to unity in comparison with $\langle \Psi | \Psi_{LS} \rangle$
in the intermediate coupling region.
In fact, for $\lambda/U$=0.1, we find
$\langle \Psi | \Psi_{j-j} \rangle$=0.972,
while $\langle \Psi | \Psi_{LS} \rangle$=0.957.

Now we discuss the results of $H_{\rm eff}$.
In Figs.~2(c) and 2(d), broken curves denote $k_4(2,4)$ and $k_6(2,4)$
evaluated from eq.~(\ref{HSeff}), respectively.
We clearly observe that the results are actually improved from the values
of the $j$-$j$ coupling limit.
In addition, it is found that they are close to the solid curves
even for $\lambda/U$ in the order of 0.1,
i.e., in the intermediate coupling region.
This fact supports the previous statement that $H_{\rm eff}$ works even
for $\lambda$ in the order of 0.1 eV, when we take $U$ in the order of eV.

In Fig.~2(f), we show the energy levels of $H_{\rm eff}$
by black solid curves for $\lambda$=0.2 eV and $U$=1 eV.
For comparison, we show the results of $H$ by red broken curves
for the same parameters.
Note that we directly diagonalize $H_{\rm eff}$ and $H$ to depict
solid and broken curves.
In the first impression, the results of $H_{\rm eff}$ are 
similar to those of $H$.
Since the effect of $B_6^0$ is now included efficiently
in the effective interaction eq.~(\ref{eq:int1}),
the CEF energy levels of $H$ are well reproduced.
In particular, we can obtain $\Gamma_{3}^{+}$ doublet ground state,
which does not appear in the limit of $\lambda$=$\infty$.

Although we do not have perfect agreements between the results of
$H_{\rm eff}$ and $H$ in Fig.~2(f),
it should be noted that we do not adjust any parameters.
In this sense, it is concluded that the characteristic features
of the CEF energy levels of $H$ are well captured by $H_{\rm eff}$.
Readers may consider that it is possible to fit the results of $H$
by the $LS$ coupling scheme due to the adjustment of the CEF parameters.
However, the adjustable parameters, $A_4$ and $A_6$ in eqs.~(\ref{eq:CEF2}),
are considered to depend on materials, not on $U$ and $\lambda$.
As remarked above, we cannot adjust the ratios of the Stevens factors,
$k_4(n,J)$ and $k_6(n,J)$, since they are determined by the symmetry
requirement and the values of $U$ and $\lambda$.
Namely, the difference in the intermediate coupling region
between the original Hamiltonian $H$ and the $LS$ coupling scheme
cannot be improved essentially,
as long as we consider the limit of $U$=$\infty$.
If we actually intend to improve the $LS$ coupling scheme,
it is necessary to consider, for instance,
the expansion in terms of $1/U$ from the limit of $U$=$\infty$.
This is an alternative way to construct the effective model
in the intermediate coupling region.

\begin{figure}[t]
\begin{center}
\includegraphics[width=8.5truecm]{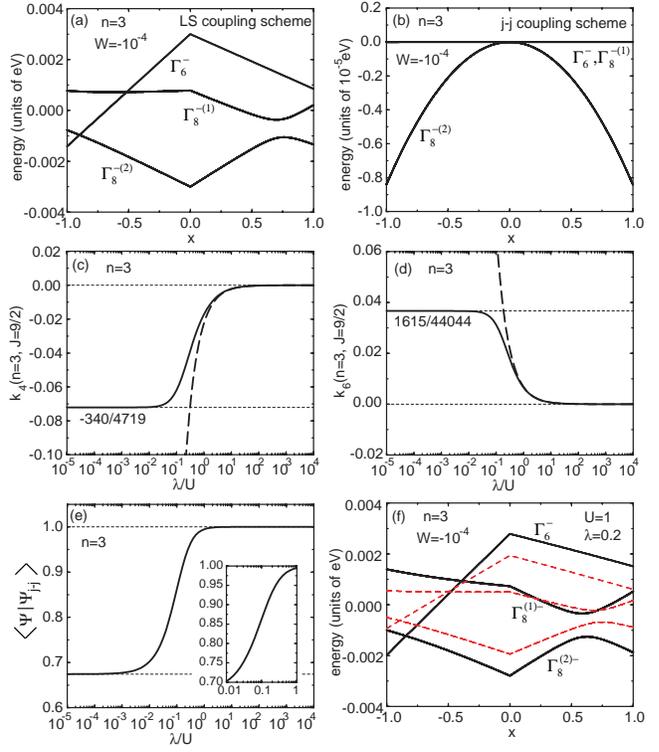}
\caption{CEF energy levels (a) in the $LS$ coupling and
(b) the $j$-$j$ coupling schemes for $n$=3.
To express virtually the $LS$ and $j$-$j$ coupling schemes, we set
($U$, $\lambda$)=($10^5$, 1) and (1, $10^5$)  for $H$
in (a) and (b), respectively.
(c) $k_4(3,9/2)$ and (d) $k_6(3,9/2)$ as functions of
$\lambda/U$ for $n$=3.
Broken curves denote $k_4(3,9/2)$ and $k_6(3,9/2)$
evaluated from $H_{\rm eff}$.
(e) Overlap integral $\langle \Psi | \Psi_{j-j} \rangle$
as functions of $\lambda/U$ for $n$=3.
Inset indicates the result in the intermediate coupling region
in a magnified scale.
(f) CEF energy levels of $H_{\rm eff}$ (black solid curves)
and $H$ (red broken curves) for $n$=3, $U$=1, and $\lambda$=0.2.}
\end{center}
\end{figure}

\subsection{$f^3$ states}

Next we consider the case of $n$=3 by following the same discussion
flow as in the previous subsection for $n$=2.
In Figs.~3(a) and 3(b), we show the CEF energy levels obtained by
the direct diagonalization of $H$ for ($U$, $\lambda$)=($10^5$, 1)
and (1,$10^5$), respectively.
In the $LS$ coupling scheme,
the ground state multiplet is characterized by $^4I_{9/2}$
($L$=6, $S$=3/2, $J$=9/2).
The 10-fold degenerate state is split into 
two $\Gamma_8^-$ and one $\Gamma_6^-$ states
under the Cubic CEF.\cite{LLW}
On the other hand, in the limit of the $j$-$j$ coupling scheme,
the situation seems quite different.
As shown in Fig.~3(b), the $\Gamma_8^-$ quartet always
becomes the ground state, while the energy for
the 6-fold degenerate excited states
including another $\Gamma_8^-$ quartet and $\Gamma_6^-$ doublet
do not depend on the CEF potential.
We find that the curve for the ground state quartet
is quadratic in terms of the CEF potential, suggesting that
the contribution in the order of $W$ vanishes in the limit of the
$j$-$j$ coupling scheme, as suggested from
$k_4^{j-j}(3,9/2)$=$k_6^{j-j}(3,9/2)$=0.
Namely, the Stevens Hamiltonian $H_{\rm S}$ does not work in this case.
However, it is found that the results of $H_a^{(0)}$ for $n$=3
correctly reproduce the CEF energy levels in Fig.~3(b).
This fact clearly indicates that it is necessary to calculate
the energy levels by diagonalizing simultaneously the Coulomb
interaction and the CEF potential terms in the $j$-$j$ coupling scheme.
As already mentioned, we always perform the direct diagonalization
of $H_{\rm eff}$ in the $j$-$j$ coupling scheme.
The Stevens Hamiltonian in the $j$-$j$ coupling scheme
is used only for the purpose to estimate $k_4(n,J)$ and $k_6(n,J)$.

Figures 3(c) and 3(d) denote $k_4(3,9/2)$ and $k_6(3,9/2)$, respectively.
We find that the values at the limits of $U$=$\infty$ and $\lambda$=$\infty$
are exactly the same as the analytic values
in the $LS$ and $j$-$j$ coupling schemes (see Table. I).
In particular, we confirm that $k^{j-j}_4(3,9/2)$=$k^{j-j}_6(3,9/2)$=0,
consistent with the quadratic behavior of the CEF energy in Fig.~3(b).
In Fig.~3(e), we depict the overlap integral as a function of $\lambda/U$.
When we decrease $\lambda/U$, the overlap integral is decreased
and at the limit of $U$=$\infty$,
we find $\langle \Psi_{LS} | \Psi_{j-j} \rangle$=0.674.
It seems to be larger than we have naively expected.
In the intermediate coupling region,
we find $\langle \Psi | \Psi_{j-j} \rangle$=0.862 for $\lambda/U$=0.1
and $\langle \Psi | \Psi_{j-j} \rangle$=0.932 for $\lambda/U$=0.2.
These results indicate that it is meaningful to construct the
$f^3$-electron state only by using the $j$=5/2 sextet
in the intermediate coupling region for $n$=3.

In Figs.~3(c) and 3(d), we also show $k_4(3,9/2)$ and $k_6(3,9/2)$
evaluated from eq.~(\ref{HSeff}) by broken curves.
They deviate from the solid curves around at $\lambda$$\sim$1,
but as shown in Fig.~3(f), even for $\lambda$=0.2 and $U$=1,
characteristic features of the CEF energy levels of $H$
are reproduced by $H_{\rm eff}$.
Thus, in the same approximation level when we use
the $LS$ coupling scheme, we can exploit $H_{\rm eff}$
for the purpose to fit the experimental results by
adjusting $A_4$ and $A_6$ in eqs.~(\ref{eq:CEF2}).
There is an advantage of $H_{\rm eff}$
that the common CEF parameters can be used,
even when the local $f$-electron number is changed.
It is concluded that our modified $j$-$j$ coupling scheme
works well to consider the multi-$f$-electron state.

\begin{figure}[t]
\begin{center}
\includegraphics[width=8.5truecm]{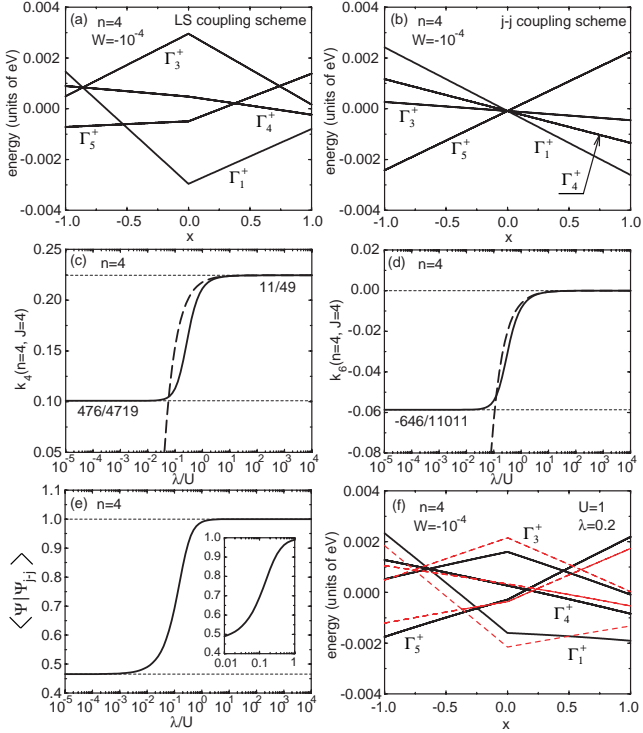}
\caption{(a) CEF energy levels in the $LS$ coupling scheme for $n$=4,
obtained by diagonalizing $H$ for $U$=$10^5$ and $\lambda$=1.
(b) CEF energy levels in the $j$-$j$ coupling scheme for $n$=4.
Here we set $U$=1 and $\lambda$=$10^5$ for $H$.
(c) $k_4(4,4)$ and (d) $k_6(4,4)$ as functions of
$\lambda/U$ for $n$=4.
Broken curves denote $k_4(4,4)$ and $k_6(4,4)$
evaluated from $H_{\rm eff}$.
(e) Overlap integral $\langle \Psi | \Psi_{j-j} \rangle$
as functions of $\lambda/U$ for $n$=4.
Inset indicates the result in the intermediate coupling region
in a magnified scale.
Inset shows the result in the intermediate coupling region.
(f) CEF energy levels of $H_{\rm eff}$ (black solid curves)
and $H$ (red broken curves) for $n$=4, $U$=1, and $\lambda$=0.2.}
\end{center}
\end{figure}

\subsection{$f^4$ states}

Encouraged by the results of $n$=2 and 3,
let us further proceed to the case of $n$=4.
In Figs.~4(a) and 4(b), we show the CEF energy levels for $n$=4
obtained by the direct diagonalization of $H$
for ($U$, $\lambda$)=($10^5$, 1) and (1,$10^5$), respectively.
For $n$=4, the ground state multiplet is characterized
by $^5I_{4}$ ($L$=6, $S$=2, $J$=4).
The nontet is split into $\Gamma_{1}^{+}$ singlet,
$\Gamma_{3}^{+}$ doublet, and two kinds of triplets
($\Gamma_{4}^{+}$ and $\Gamma_{5}^{+}$).\cite{LLW}
Since the value of $J$ for $n$=4 is equal to that of $n$=2,
the CEF states are in common with Fig.~2(a), but
the $x$ dependence is different, since the sign of
$k^{LS}_4(4,4)$ is different from $k_4^{LS}(2,4)$.
Due to the same reason, we find that the $x$ dependence
of the CEF energy levels in Fig.~4(b) is reversed from
that of Fig.~2(b) in the $j$-$j$ coupling scheme.

In Figs.~4(c) and 4(d), we show $k_4(4,4)$ and $k_6(4,4)$ by solid curves,
indicating that the values in the $LS$ and the $j$-$j$ coupling limits
are correctly reproduced with the smooth changes between them.
When we compare them with the results for $k_4(4,4)$ and $k_6(4,4)$
evaluated from eq.~(\ref{HSeff}) (broken curves),
they begin to deviate from the solid curves around at $\lambda$=2$\sim$3.
As shown in Fig.~4(e), at the limit of $U$=$\infty$,
we obtain $\langle \Psi_{LS} | \Psi_{j-j} \rangle$=0.465 for $n$=4.
In the intermediate coupling region,
we find $\langle \Psi | \Psi_{j-j} \rangle$=0.692 for $\lambda/U$=0.1
and $\langle \Psi | \Psi_{j-j} \rangle$=0.836 for $\lambda/U$=0.2.
Even for $n$=4, in the intermediate coupling scheme,
the ground state is well approximated by the $j$-$j$ coupling scheme.
Then, it is still possible to
reproduce the results in the intermediate coupling region.
In fact, as shown in Fig.~4(f), the CEF energy levels of $H$
look similar to those of $H_{\rm eff}$ for $U$=1 and $\lambda$=0.2.

\subsection{$f^5$ states}

Now we move on to the case of $n$=5.
In Figs.~5(a) and 5(b), we depict the CEF energy levels due to
the direct diagonalization of $H$ with $n$=5
for ($U$, $\lambda$)=($10^5$, 1) and (1,$10^5$), respectively.
For $n$=5, the ground state multiplet is characterized
by $^6H_{5/2}$ ($L$=3, $S$=5/2, $J$=5/2).
The sextet is split into $\Gamma_{7}^{-}$ doublet
and $\Gamma_{8}^{-}$ quartet.\cite{LLW}

Here readers may consider that the CEF energy levels for $n$=5
are simply the same as those in the case of $n$=1,
but we should not simply conclude it.
Namely, as easily understood from Figs.~5(a) and 5(b),
the $x$ dependence of the CEF energy levels is just reversed
between the $LS$ and $j$-$j$ coupling schemes.
It should be noted that this phenomenon is $not$ due to
the approximation, but the intrinsic feature of
the original Hamiltonian $H$.
Note also that for $n$=1 and 5, the only relevant CEF parameter
is $B_4(n,J)$, since the sixth-order CEF potential terms do not
appear in the space of $J$=5/2.
In the limit of the $LS$ coupling scheme,
since $k_4(5,5/2)$ and $k_4(1,5/2)$ have the same sign,
the ground state character should not be changed between
the cases of $n$=1 and 5.

On the other hand, in the limit of the $j$-$j$ coupling scheme,
the $x$ dependence of the CEF ground state of $n$=5 is reversed
from that of $n$=1, since the sign of $k_4(5,5/2)$ is different
from that of $k_4(1,5/2)$ in the $j$-$j$ coupling limit,
as shown in Table I.
We can understand the reason more intuitively
from the electron-hole relation on the basis of the $j$-$j$
coupling scheme.
Namely, to obtain the $f^5$-electron state in the $j$-$j$ coupling scheme,
we accommodate five electrons in the one-electron potential levels,
for instance, such as $\Gamma_7^-$ ground and $\Gamma_8^-$ excited states
for $B_4^0(1,5/2)$$>$0 ($x$$<$0 and $W$$<$0).
Then, we obtain $\Gamma_8^-$ ground and $\Gamma_7^-$ excited states
for the case of $n$=5 in the same region of $x$.
In Fig.~5(c), we depict $k_4(5,5/2)$ as a function of $\lambda/U$ for $n$=5.
We confirm that $k_4(5,5/2)$ smoothly changes
from $-11/7$ at $\lambda$=$\infty$ to $13/21$ at $U$=$\infty$.

\begin{figure}[t]
\begin{center}
\includegraphics[width=8.5truecm]{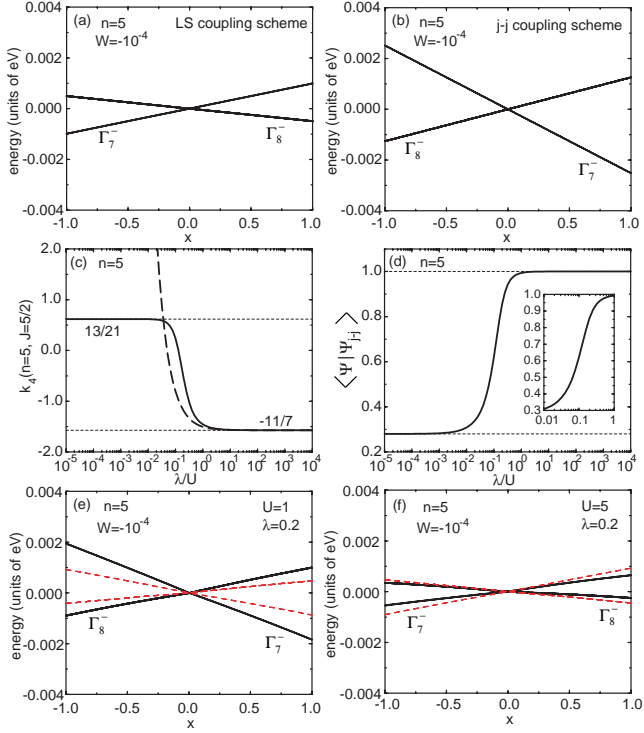}
\caption{(a) CEF energy levels in the $LS$ coupling scheme for $n$=5.
We diagonalize $H$ for $U$=$10^5$ and $\lambda$=1.
(b) CEF energy levels of $H$ in the $j$-$j$ coupling scheme for $n$=5.
Here we set $U$=1 and $\lambda$=$10^5$.
(c) $k_4(5,5/2)$ as a function of $\lambda/U$ for $n$=5.
Broken curve denotes the result estimated from $H_{\rm eff}$.
(d) Overlap integral $\langle \Psi | \Psi_{j-j} \rangle$
as a function of $\lambda/U$ for $n$=5.
Inset denotes the result in the intermediate coupling region
in a magnified scale.
CEF energy levels of $H_{\rm eff}$ and $H$
for (e) $U$=1 and (f) $U$=5.5 with $n$=5 and $\lambda$=0.2.
In both figures, we depict the results of $H_{\rm eff}$
and $H$ by black solid and red broken curves, respectively.}
\end{center}
\end{figure}

When we compare the solid curves with the broken ones
for $k_4(5,5/2)$ estimated from eq.~(\ref{HSeff}),
the deviation begins to occur at $\lambda$ between 1 and 10.
Namely, the degree of approximation seems to be worse
in comparison with the case of $n$=4.
In fact, we obtain the smaller overlap integral for $n$=5,
$\langle \Psi_{LS} | \Psi_{j-j} \rangle$=0.280,
as shown in Fig.~5(d).
However, in the intermediate coupling region,
we find $\langle \Psi | \Psi_{j-j} \rangle$=0.622 at $\lambda$=0.1
and $\langle \Psi | \Psi_{j-j} \rangle$=0.857 at $\lambda$=0.2,
for a fixed value of $U$=1.
At $\lambda$=0.2, $\langle \Psi | \Psi_{j-j} \rangle$ for $n$=5
is larger than that for $n$=4, as discussed in Sec.~2.2.
In any case, the ground-state wavefunction in the intermediate
coupling region is still approximated well by the states constructed
from the $j$-$j$ coupling scheme for $n$=5.
We can reproduce the results of $H$ by $H_{\rm eff}$
for $U$=1 and $U$=5 at $\lambda$=0.2,
as shown in Figs.~5(e) and 5(f).
In particular, in the effective model of $H_{\rm eff}$,
we can reproduce the interchange of the ground states,
when we change the ratio of $\lambda/U$.
Note, however, that we cannot quantitatively reproduce
the critical value of $U$ at which the ground states
are interchanged.

In order to understand the interchange of the ground state
for the case of $n$=5 within the $LS$ coupling scheme,
the only way is to change the sign of $W$ phenomenologically,
since $k_4^{LS}(5,5/2)$ is fixed due to the symmetry requirement.
However, in our modified $j$-$j$ coupling scheme, we can consider
the microscopic origin of the change of
the CEF parameter due to the competition between
the Coulomb interaction and the spin-orbit coupling.
It is an advantage of our effective model,
in addition to the reproduction of the CEF energy levels of $H$.

As for the change of the sign in the CEF parameter,
when we consider the different material groups, we may choose
positive or negative $W$ depending on the kind of materials.
However, for the same material group with the same rare-earth ion,
it is difficult to imagine that the sign of the CEF parameter is changed,
although the magnitude may be different due to the change of ligand ions.
Rather, as found in the present calculations, it seems natural
to understand that the CEF ground state is converted due to
the effect of the Coulomb interaction and/or the spin-orbit coupling.
This is considered to explain a possible conversion of the CEF ground state
in Sm-based filled skutterudites, as mentioned in the next subsection.

\subsection{$T_{\rm h}$ symmetry}

In order to show the effectiveness of our modified $j$-$j$
coupling scheme, let us also discuss the CEF energy levels for
filled skutterudites with $T_{\rm h}$ symmetry.\cite{Sato,Aoki}
For the purpose, it is necessary to add extra
$B_6^2$ terms~\cite{Takegahara} in $B_{m,m'}$ as~\cite{Hotta3}
\begin{eqnarray}
  \begin{array}{l}
    B_{3,1}=B_{-3,-1}=24\sqrt{15}B_6^2,\\
    B_{2,0}=B_{-2,0}=-48\sqrt{15}B_6^2,\\
    B_{1,-1}=-B_{3,-3}=360B_6^2.
  \end{array}
\end{eqnarray}
By following Ref.~\citen{Takegahara}, we express $B_6^2$ as
$B_6^2$=$Wy/F^t(6)$ with $F^t(6)$=24 for $J$=3.
Here we set $y$=0.3
as a typical value for filled skutterudites.

In Figs.~6(a) and 6(b), we show the energy levels for $n$=1 and 2
by diagonalizing $H_{\rm eff}$ for $\lambda$=0.1 and 0.2, respectively,
with $U$=1.
For the case of $n$=2,
we also depict the results of $H$ by red broken curves.
For $n$=1, we have $\Gamma_{5}^{-}$ doublet and $\Gamma_{67}^{-}$ quartet
states, which are essentially the same as $\Gamma_{7}^{-}$ doublet and
$\Gamma_{8}^{-}$ quartet for $O_{\rm h}$ symmetry.
For $n$=2, we find remarkable difference from the case of
$O_{\rm h}$ symmetry.
As already mentioned in Ref.~\citen{Takegahara}, two triplets
$\Gamma_{4}^{+}$ and $\Gamma_{5}^{+}$ in $O_{\rm h}$ symmetry
are mixed and they are $\Gamma_{4}^{+(1)}$ and $\Gamma_{4}^{+(2)}$
in $T_{\rm h}$ symmetry.
It is found that the results of $H$ in the intermediate coupling
region are well reproduced by $H_{\rm eff}$.
It is quite natural, since the effect of $B_6^2$ is effectively
included in $H_{\rm eff}$.
It is emphasized here that such a characteristic issue of
$T_{\rm h}$ symmetry is correctly reproduced
in our modified $j$-$j$ coupling scheme.

\begin{figure}[t]
\begin{center}
\includegraphics[width=8.5truecm]{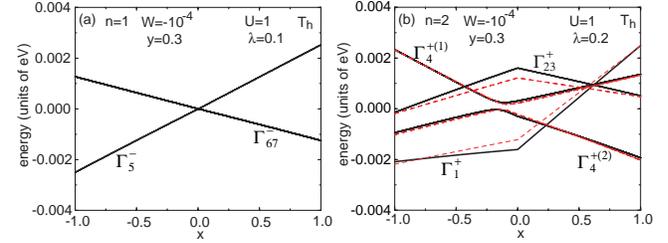}
\caption{
CEF energy levels of $H_{\rm eff}$ (black solid curves) for
(a) $n$=1 and (b) $n$=2 for $T_{\rm h}$ symmetry with $y$=0.3.
The spin-orbit interaction is set as (a) $\lambda$=0.1 and
(b) $\lambda$=0.2.
In the panel (b), we also show the results of $H$
by red broken curves for the same parameters.}
\end{center}
\end{figure}

For $n$=5, we have $\Gamma_{5}^{-}$ ($\Gamma_{7}^{-}$ in $O_{\rm h}$)
doublet and $\Gamma_{67}^{-}$ ($\Gamma_{8}^{-}$ in $O_{\rm h}$)
quartet states.
Thus, the results are essentially the same as those
in Figs.~5(e) and 5(f).
As mentioned above, for Sm-based filled skutterudites,
it has been recently pointed out a possibility
that the ground states are changed between
$\Gamma_{5}^{-}$ doublet and $\Gamma_{67}^{-}$ quartet.
For SmRu$_4$P$_{12}$, SmOs$_4$P$_{12}$, and SmOs$_4$Sb$_{12}$,
it has been considered that $\Gamma_{67}^-$ quartet is the ground state.
\cite{Sanada,Matsuhira,Aoki-Sm}
On the other hand, for SmFe$_4$P$_{12}$, a possibility of
$\Gamma_5^-$ doublet ground state has been suggested.
\cite{Matsuhira,Nakanishi-Sm}
It can be interpreted that the conversion of the CEF ground state
is due to the change of the Coulomb interaction and/or
the spin-orbit coupling, which naturally leads to the change in
the sign of $B_4^0$.
This point will be discussed in detail elsewhere with numerical results
on the possible multipole state of Sm-based filled skutterudites.
\cite{Hotta-Sm}

%
%
\section{Discussion and Summary}

In this paper, we have proposed the effective model
to consider systematically the multi-$f$-electron states
by using the expansion in terms of $1/\lambda$.
We have shown that the results in the intermediate coupling region
for  $n$=2$\sim$5 can be reproduced well by the effective model
and the applicability of the model has been clarified.
The CEF state for $T_{\rm h}$ symmetry can be also reproduced.

We have stated that $H^{(0)}_a$ cannot reproduce
even qualitatively the energy levels of the intermediate coupling
region as well as the $LS$ coupling scheme,
since the effect of $B_6^0$ is not included.
Here readers may have a question:
If the effect of $B_6^0$ is included in the one-electron potential,
does the zeroth-order term mimic the $LS$ coupling results?
The answer is yes.
This can be clarified by considering the cases of $n$=13 and 12,
i.e., the situations of one and two holes in the $j$=7/2 octet.
For $j$=7/2, the zeroth-order term of $1/\lambda$ is given by
\begin{equation}
  \begin{split}
   H^{(0)}_{b} & = \sum_{\beta_1,\beta_2}
   {\tilde B}^{b,b}_{\beta_1,\beta_2}
   f^{\dag}_{b\beta_1}f_{b\beta_2} \\
   &+ \sum_{\beta_1 \sim \beta_4}
   {\tilde I}^{bb,bb}_{\beta_1\beta_2,\beta_3\beta_4}
   f^{\dag}_{b\beta_1}f^{\dag}_{b\beta_2}f_{b\beta_3}f_{b\beta_4},
  \end{split}
\end{equation}
where ${\tilde B}^{b,b}_{\beta_1,\beta_2}$ is given by
the CEF potential for $j$=7/2
and ${\tilde I}^{bb,bb}_{\beta_1\beta_2,\beta_3\beta_4}$
is the Coulomb interaction among $j$=7/2 octet.

\begin{figure}[t]
\begin{center}
\includegraphics[width=8.5truecm]{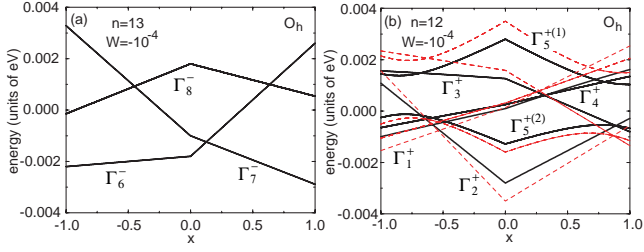}
\caption{CEF energy levels of $H_b^{(0)}$ (black solid curves)
for (a) $n$=13 and (b) $n$=12 with $U$=1 and $W$=$-10^{-4}$.
For comparison, in the panel (b),
we also show the results of $H$ in the $LS$ coupling scheme
($U$=$10^5$ and $\lambda$=1) by red broken curves.}
\end{center}
\end{figure}

In Figs.~7(a) and 7(b), we show the energy levels of $H^{(0)}_b$
by black solid curves for $n_b$=7 and 6,
corresponding to $n$=13 and 12, respectively,
where $n_b$=$n-6$ and $n_b$ denotes the electron number in the $j$=7/2 octet.
Note that the results are obtained by the diagonalization of $H_b^{(0)}$.
For comparison, for the case of $n$=12,
we also show the results in the $LS$ coupling scheme,
by setting $U$=$10^5$ and $\lambda$=1 for $H$.
In the $j$=7/2 octet, the effect of $B_6^0$ is already included
in the one-hole case ($n$=13),
as easily understood from the Hutchings table for $J$=7/2.
For the two-hole case ($n$=12), as shown in Fig.~7(b),
even without considering the $1/\lambda$-correction term $H_b^{(1)}$,
the CEF energy levels of $H^{(0)}_{b}$ (black solid curves)
are quite similar to those in the $LS$ coupling scheme for $J$=6.
Of course, in order to reproduce quantitatively the CEF energy levels
in the realistic intermediate coupling region, it is necessary
to include the terms in the order of $1/\lambda$ for the $j$=7/2 octet,
as has done in this paper for the $j$=5/2 sextet.
However, for practical purposes
such as the fitting of experimental results,
it seems enough to use $H^{(0)}_{b}$ for Tm and Yb compounds.
When we further proceed to the cases of $n$$<$12,
it is recommended to improve $H^{(0)}_{b}$
by considering the $1/\lambda$ corrections.
It is one of future issues, when we attempt to develop a microscopic
theory of magnetism in heavy rare-earth compounds.


As for a possible application of our modified $j$-$j$ coupling scheme,
we consider a direction to improve effectively
the band-structure calculations for $f$-electron materials.
In the relativistic band-structure calculations,
all the one-electron CEF potentials for $j$=5/2 and 7/2 states
are correctly included,\cite{Harima}
but the multi-$f$-electron state is not completely reproduced
in comparison with the CEF levels of the $LS$ coupling scheme.
In order to construct the multi-$f$-electron state
due to the one-electron basis,
it is necessary to treat correctly the competition among
Coulomb interactions, spin-orbit coupling, and CEF potentials,
but in the band-structure calculations,
the effect of Coulomb interactions, in particular,
the Hund's-rule coupling, is considered only partly
within the mean-field approximation.
An effective way to improve such a situation is
to include the two-body CEF potentials discussed here
in the band-structure calculations with due care.
It is one of future problems to develop a systematic way
for the inclusion.

Another issue is the CEF effect on exotic itinerant magnetism
and unconventional superconductivity of $f$-electron materials
from the band picture.
We can construct the appropriate many-body Hamiltonian
for $f$-electron systems,
by adding the hybridization term between $f$ and conduction electrons
or the hopping term of $f$ electrons to the local $f$-electron term
$H_{\rm eff}$ in the modified $j$-$j$ coupling scheme.
The parameters in the kinetic term are determined so as to reproduce
the energy bands near the Fermi level.
It is another future problem to analyze such a model by using
numerical and/or analytical techniques.


In summary, we have discussed the $f^n$-electron states
with $n$$\ge$2 on the basis of the effective model obtained
by including the corrections in the order of $1/\lambda$
in the $j$-$j$ coupling scheme,
For $n$=2$\sim$5, the results in the realistic intermediate
coupling region have been quantitatively reproduced
in our modified $j$-$j$ coupling scheme.
By using $H_{\rm eff}$, we have also reproduced correctly
the CEF energy levels for $T_{\rm h}$ symmetry.
In conclusion, for the consideration of multi-$f$-electron states,
we can use the $j$-$j$ coupling scheme with appropriate corrections
in terms of $1/\lambda$.

\section*{Acknowledgment}

The authors thank K. Kubo, H. Onishi, and K. Ueda
for discussions and comments.
This work was supported by a Grant-in-Aid for Scientific Research (C)(2)
under the contract No.~16540316
from Japan Society for the Promotion of Science (JSPS).
We have been separately supported by Grant-in-Aids for
Scientific Research in Priority Area ``Skutterudites''
under the contract Nos.~18027016 and 15072204
from the Ministry of Education, Culture, Sports, Science,
and Technology of Japan.
One of the authors (T.H.) has been also supported by a Grant-in-Aid for
Scientific Research (C) under the contract No.~18540361 from JSPS.
Part of the computation in this work has been done using the facilities
of the Supercomputer Center of Institute for Solid State Physics,
University of Tokyo.


\end{document}